# Negative oil price bubble
# is likely to burst in March – May 2016.
# A forecast on the basis of
# the law of log-periodical dynamics

***Fomin, Alexey A.***, Center for Log-Term Forecasting and Strategic Planning at the Moscow State University

***Korotayev, Andrey V.***, Laboratory for Monitoring of Sociopolitical Destabilization Risks of the National Research University Higher School of Economics, Moscow; System Forecasting Center of Oriental Institute and Institute for African Studies, Russian Academy of Sciences, Moscow

***Zinkina, Julia V.***, Laboratory for Monitoring of Sociopolitical Destabilization Risks of the National Research University Higher School of Economics, Moscow; Institute for African Studies, Russian Academy of Sciences, Moscow

Data analysis with log-periodical parametrization of the Brent oil price dynamics has allowed to estimate (very approximately) the date when the dashing collapse of the Brent oil price will achieve the absolute minimum level (corresponding to the so-called singularity point), after which there will occur a rather rapid rebound, whereas the accelerating fall of the oil prices which started in mid-2014 will come to an end. This is likely to happen in the period between March, 24$^{th}$ and May, 15$^{th}$, 2016. An analogous estimate (though a more exact one) was made for the date of the burst of the nearest "sub-bubble", which is likely to occur between 19.01 and 02.02.2016 (importantly, this estimate will allow to verify the robustness of the developed forecast in the very nearest days). However, this will not mean a start of a new uninterrupted global growth – the fall will soon continue, breaking new "anti-records". The fall will only finally stop after passing the abovementioned point of the main negative bubble singularity somewhere between March 24$^{th}$ and May 15$^{th}$, 2016. Importantly, our calculations have also shown that after mid-2014 we are dealing not with an antibubble (when price collapse goes on in a damped and almost unstoppable regime) in the world oil market, but with a negative bubble, when prices collapse in an accelerated mode, and there can be particularly powerful collapses with particularly strong destabilizing effect near the singularity point. On the other hand, negative bubbles can be better manipulated by the actions of the macro actors.

**Introduction**

Data analysis with log-periodical parametrization of the Brent oil price dynamics has allowed to estimate (very approximately) the date when the dashing collapse of the Brent oil price will achieve the absolute minimum level (corresponding to the so-called singularity point), after which there will occur a rather rapid rebound, whereas the accelerating fall of the oil prices which started in mid-2014 will come to an end. This is likely happen in the period between March, 24[th] and May, 15[th], 2016[1].

An analogous estimate (though a more exact one) was made for the date of the burst of the nearest "sub-bubble", which is likely to occur between 19.01 and 02.02.2016 (importantly, this estimate will allow to verify the robustness of the developed forecast in the very nearest days). However, this will not mean a start of a new uninterrupted global growth – the fall will soon continue, breaking new "anti-records". The fall will only finally stop after passing the abovementioned point of singularity somewhere between March 24[th] and May 15[th], 2016.

Importantly, our calculations have also shown that after mid-2014 we are dealing not with an antibubble (when price collapse goes on in a damped and almost unstoppable regime) in the world oil market, but with a negative bubble, when prices collapse in an accelerated mode, and there can be particularly powerful collapses with particularly strong destabilizing effect near the singularity point. On the other hand, negative bubbles can be better manipulated by the actions of the macro actors.

**Log-periodic parametrization**

In a number of seminal works by Didier Sornette, Anders Johansen and their colleagues (Sornette 2004; Sornette, Johansen 1997, 1998, 2001; Sornette, Sammis 1995; Sornette, Woodard, Zhou 2009; Johansen, Sornette 1999, 2001; Johansen, Sornette, Ledoit 1999; Johansen *et al.* 1996) it has been demonstrated that accelerating log-periodic oscillations superimposed over an explosive growth trend that is described with a power-law function with a singularity (or quasi-singularity) in a finite moment of time $t_C$, are observed in situations leading to crashes and catastrophes. They can be analyzed as their precursors which allow the forecasting of such events. One can mention such examples as the log-periodic oscillations of the Dow Jones Industrial Average (DJIA) that preceded the crash of 1929 (Sornette, Johansen 1997), or the changes in the ion concentrations in the underground waters that preceded the catastrophic Kobe earthquake in Japan on the 17th of January, 1995 (Johansen et al. 1996), which are also described mathematically rather well with log-periodic fluctuations superimposed over a power-law growth trend.

As has been shown by these studies, what is referred to as a bubble has a clear mathematical expression:

$$p(t) = p_{max} - C_1 (t_c - t)^{C_3} \{1 + C_2 \cos[C_4 \ln(t_c - t) - \varphi]\},$$
$$\text{or} \qquad (1)$$
$$p(t) = A + m (t_c - t)^{\alpha} \{ 1 + C \cos[\omega \ln(t_c - t) + \varphi] \}.$$

---

[1] If, of course, the oil market remains at the disposal of speculators, and no massive interventions of macro-actors are made.

Here *p(t)* is some indicator of financial (or commodity) markets (for example, the price of oil). In the case of a sufficiently long period of time a more accurate result can be obtained by using the logarithm of the price (or other indicators) instead of the price itself.

*A* (or $p_{max}$) is the maximum (or, as we shall see below, the minimum) value that the corresponding indicator could reach at the critical point in time $t_c$, hereinafter referred to as the point of quasi-singularity. The role of such an indicator can be played by dollar in relation to any currency or commodity (e.g., gold or oil). This name is used because as you get closer to this point the oscillation frequency (described by a harmonic function of the logarithm) tends to infinity. At the same time, although the function *p(t)* does not tend to infinity, it makes a "leap". It does not tend to infinity due to the fact that the constant *m* is negative, while the constant *α* is positive. These oscillations are called log-periodic oscillations, because if these fluctuations are viewed in the logarithmic scale with the starting point at the point of singularity ($t_c$), they look like periodic (example can be found in Fig. 13). A feature of the log-periodic oscillations is that the period of each subsequent log-oscillation decreases as compared to the period of the previous one by the same rate equal to exp ($2π / ω$).

From a practical point of view the pattern discovered by Sornette is important because it allows to estimate the singularity point. After this point the log-periodic law stops working. And, thus, it allows to estimate the time when the growth or decline of a particular index/price stops.

**Major negative bubble**

Importantly, in the case of oil prices, one can estimate the time when the oil price collapse (going on since May 2011 and particularly strong since mid-2014) stops (Fig. 1). Defining the time of collapse is important for the oil-producing countries, because after that the global oil prices will start growing.

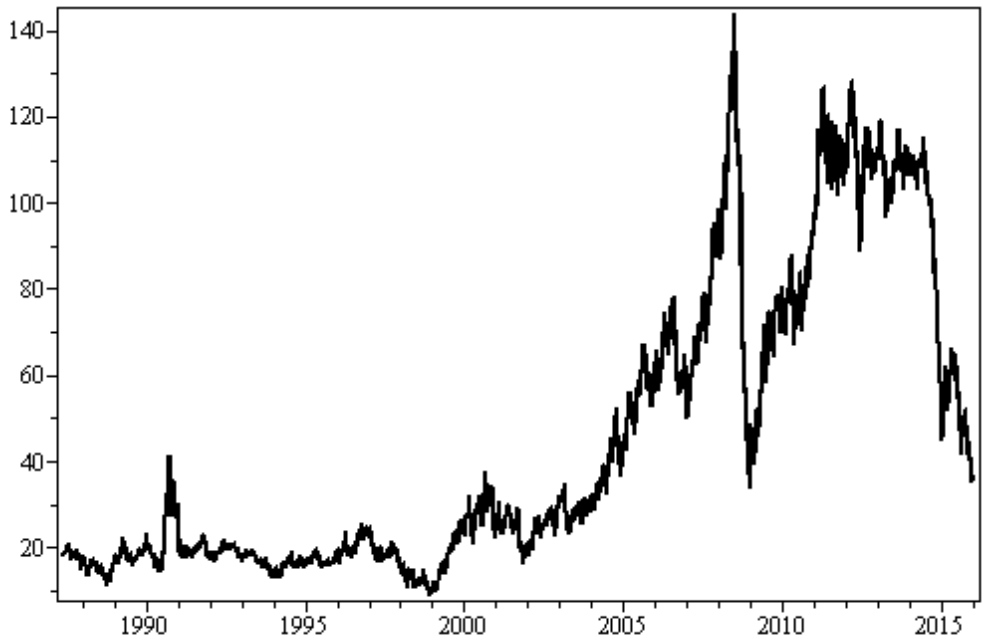

**Fig. 1. Brent oil price dynamics.** Source: *Eia.gov.* X-axis: *20.05.1987 — 04.01.2016.*

Regularity (1) makes it possible to describe this price collapse dynamics. In this case we deal with a so-called "negative bubble" – when a regular decrease of the period of log-oscillations is superimposed over a power-law *decline* trend[2] with the overall pattern described by equation (1) (see Yan *et al.* 2011).

In the dynamics of oil prices in Fig. 2 one can see three oscillation periods with declining periods. Unfortunately, the method does not allow to correctly determine the minimum price level, which may occur at the bubble collapse point.

However, the point of singularity in Fig. 2 suggests an estimate of the earliest time when the global rising trend in oil prices could probably begin.

This does not mean that after that there will be no dramatic collapses. But if three waves of negative bubble collapse that can be seen in Fig. 2 and that correspond to three log-waves occurred against the background of a general price decline, after the point of singularity collapses will be occurring against the background of overall growth.

In addition, the abovementioned estimate has its own error. In this study we do not provide a detailed estimate of this error. Judging by the error in a "small" negative sub-bubble, which will be discussed below, the distance between the error margins for the point of singularity of the negative bubble in Fig. 2 may be up to two months.[3] Thus, it is most likely that in the first half of 2016 the global decline in oil prices will stop.

Before this happens, we are likely to see several slowdowns in the decline in oil prices, and even some growth, followed by a sharp fall – that will correspond to a number of periods of log-oscillations. Data points in Fig. 2 indicate only three such oscillations. But there will be more similar oscillations with constantly declining periods.

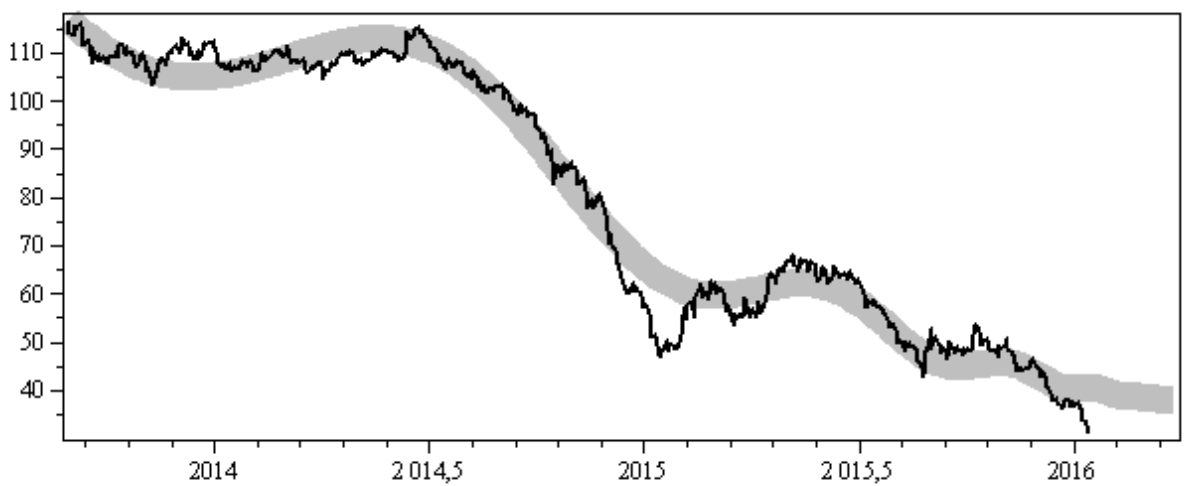

**Fig. 2. Major negative oil price bubble.** *Broken line depicts the empirical Brent oil price dynamics (source: Quandl.com). Smooth line corresponds to the following log-periodic parametrization $37.766 + 25.874 (2016.233-t)^{1.476} + 6.953 (2016.233-t)^{1.476} cos(18.950$*

---

[2] Positive bubbles are also characterized by a regular decrease of the period of log-oscillations, but in the case of positive bubbles is superimposed over a power-law *growth* trend (see, *e.g.*, Sornette 2004).

[3] Below we estimate the error of the point of singularity of the negative sub-bubble at about 0.5 months; however, it lasts a lot less than the major negative bubble (Fig. 2). In view of the fact that the price structure has a fractal nature (larger negative bubbles are somewhat similar to smaller sub-bubbles) we can make a rough estimate of the error in the point of singularity (Fig. 2). Our preliminary calculations along the lines described below for the sub-bubble indicate the interval between 24.03.2016 and 15.05.2016.

*log(2016.233-t)+2.113). Optimization interval (where the broken line lies): 12.02.2014 - 03.10.2011. Point of singularity = 2016,23279 = 25.03.2016 0:38. Smooth line is continued into the future according to the abovementioned formula until the point of singularity.*

## "Small" negative (sub) bubble

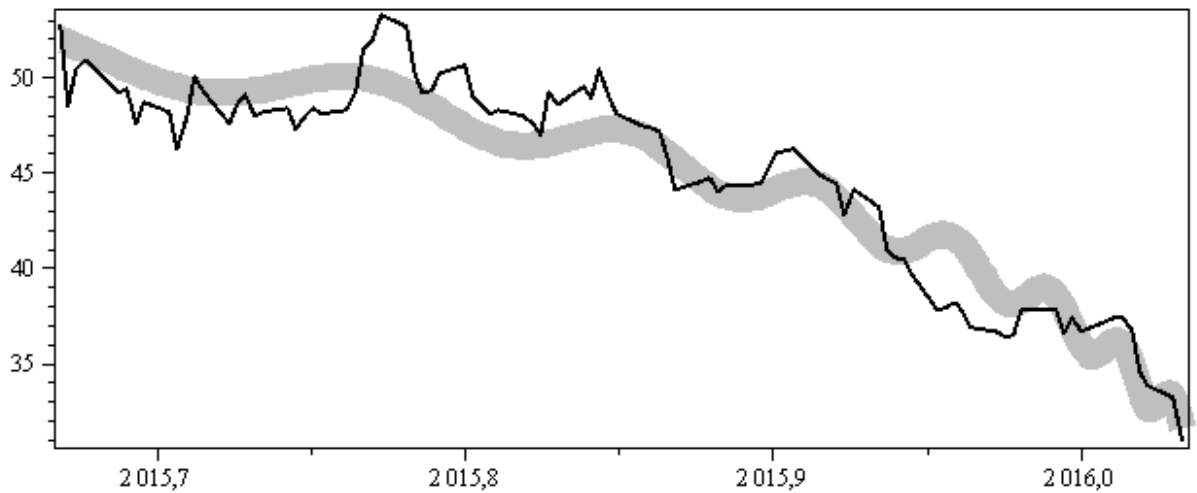

**Fig. 3. Minor negative oil price sub-bubble.** *Broken line depicts the empirical Brent oil price dynamics (source: Quandl.com). Smooth line corresponds to the following log-periodic parametrization: -35116.565 + 35175.496 $(2016.072-t)^{0.000240}$ + 1.037 $(2016.072-t)^{0.000240}$ cos{44.131 log(2016.072-t)+2.298}. Optimization interval: 15.02.2012 — 12.01.2016. Singularity point: 2016.0719 = 26.01.2016 6:17.*

Similarly to the forecasting of the date before which one should not expect[4] the overall end of the oil price decline corresponding to the major oil price bubble of Fig. 2, it appears possible to forecast the date in whose area one should expect the end of the price decline corresponding to the 3$^{rd}$ (right-hand) wave of the major bubble of Fig. 2 – as at this wave one can observe the formation of a smaller negative bubble that we will denote below as a negative sub-bubble.

---

[4] If, of course, the oil market remains at the disposal of speculators, and no massive interventions of macro-actors are made.

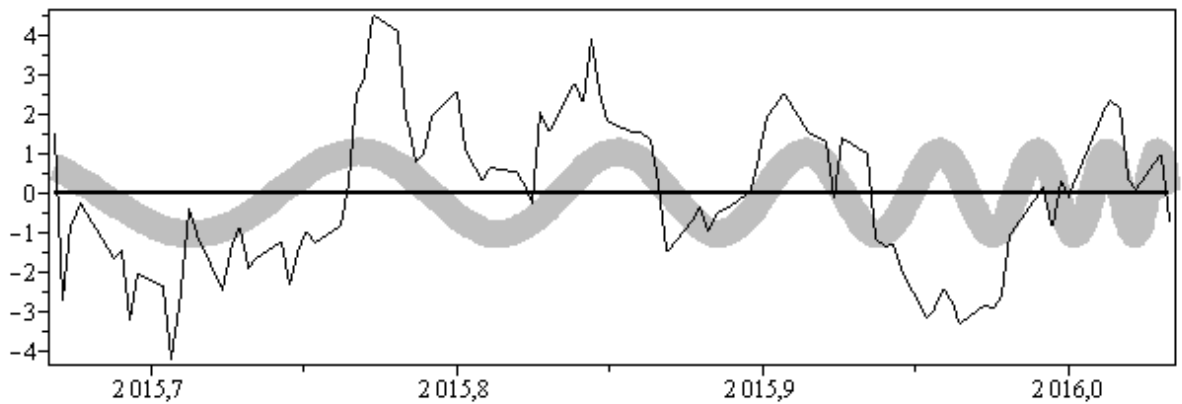

**Fig. 3. Oscillatory component of the "small" negative bubble.** *Broken line depicts the difference between the broken line of Fig. 3 and smooth (power-law) component of its parametrization (described with the following equation: -35116,565 + 35175,496 (2016,072-t)$^{0,000240}$), the smooth line corresponds to the oscillatory component of Fig. 3 (described with the following equation: 1.037 (2016.072-t)$^{0.000240}$ cos{44.131 log(2016.072-t)+2.298}.*

Here we will provide a detailed description of the methodology of estimation of the precision of the calculation of the respective singularity point.

The (sub)bubble corresponding to the 3$^{rd}$ wave of Fig. 2 is not as clear as the bubble of Fig. 2 and is depicted in Fig. 3.

The oscillatory component of the previous figure is shown in Fig. 4. It will now be used for the visual assessment of the error point of singularity. Below we will also show that – in spite of the blurred parameterization – we are dealing with a real negative bubble.

## Extrapolations into the future

Unfortunately, the extrapolation of this parametrization into the future until the singularity point of the negative sub-bubble in question demonstrate its senselessness, as it goes into the zone of negative values (Fig. 5). Therefore, the main practical value of parameterization can consist only in determining the singular point itself, after which the negative sub-bubble will not be able to develop further.

The point of singularity detected in Fig. 3, has its own error. Therefore, it cannot be regarded as a guaranteed date after which the corresponding wave of price collapse will stop.

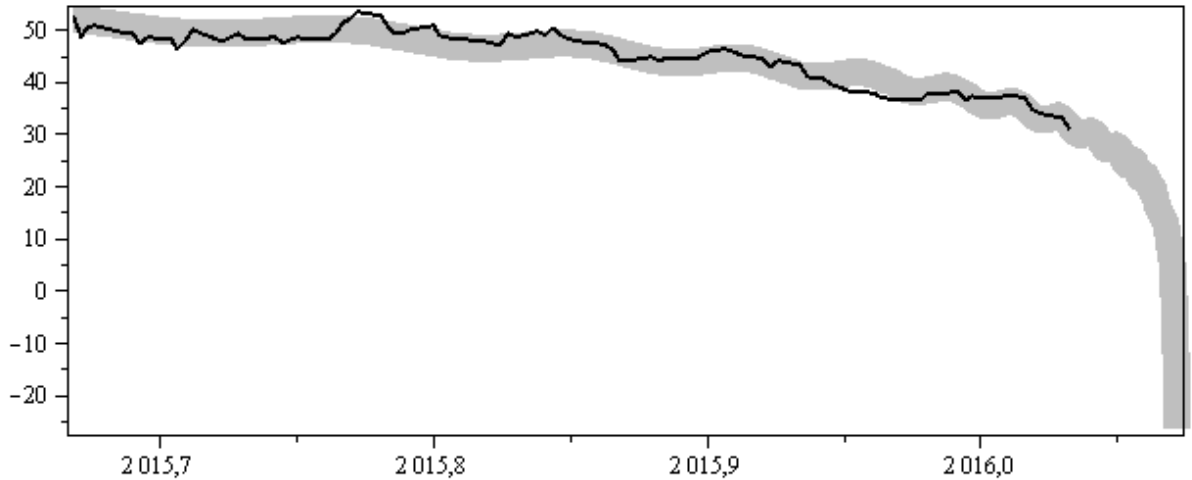

**Fig. 4. Extrapolation of the minor sub-bubble into the future does not have a predictive force.** *The same as in Fig. 3, but with the smooth line extended down to the singularity point of Fig. 3. X-axis: 26.01.2016 6:16 — 26.01.2016 6:16.*

## Corroboration of the point that we are dealing with a negative bubble

However, before estimating the error of the singularity point detected above, in this section we will present evidence that in Fig. 3 we deal with a negative bubble, and not with an anti-bubble.

Parametrization with anti-bubble

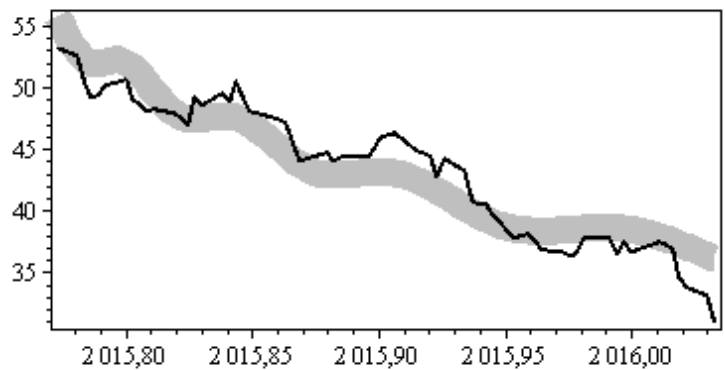

**Fig. 5. The best-fit anti-bubble.** *Smooth line corresponds to the following log-periodic parametrization: $41896966.51 - 41896944.03\ (2015.7-t)^{0.00000296} + 1.028\ (2015.7-t)^{0.00000296} \cos\{39.533 \log(2015.7-t) + 1.520\}$. Singularity point has been taken without optimization.*

In this subsection we consider an alternative parametrization of the 3$^{rd}$ wave of Fig. 1 as an antibubble. We show that it does not work, suggesting that in Fig. 3 we are dealing with a negative sub-bubble.

At first glance at Fig. 6, the idea that we are dealing here with an antibubble does not look absurd (let us recollect that the antibubble is described with the same basic equation (1) as the negative bubble is, but, in contrast with the negative bubble, with the antibubble the period of oscillations is regularly increasing [see Sornette 2004]). However, Fig. 7 indicates that we are dealing with a negative bubble rather than an antibubble.

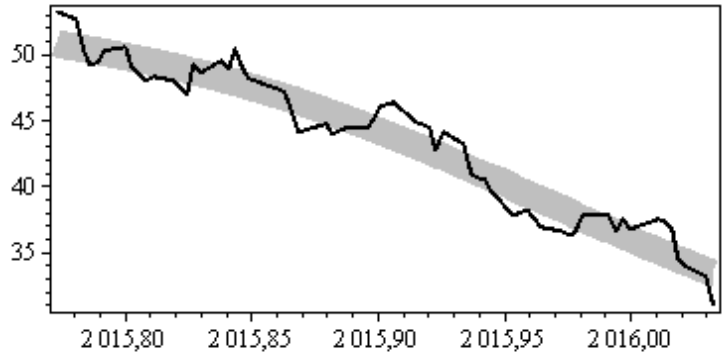

**Fig. 6. Anti-bubble turns out to be degenerate**

Fig. 6 has not been optimized for the point of singularity intentionally. Because if this is done, the situation where the extremes of the smooth line roughly match those for the broken line is impossible: if you make parameterization for the point of singularity as well, you get, instead of Fig. 6, a degenerate result shown in Fig. 7.

**A larger alternative of a "smaller" negative bubble**

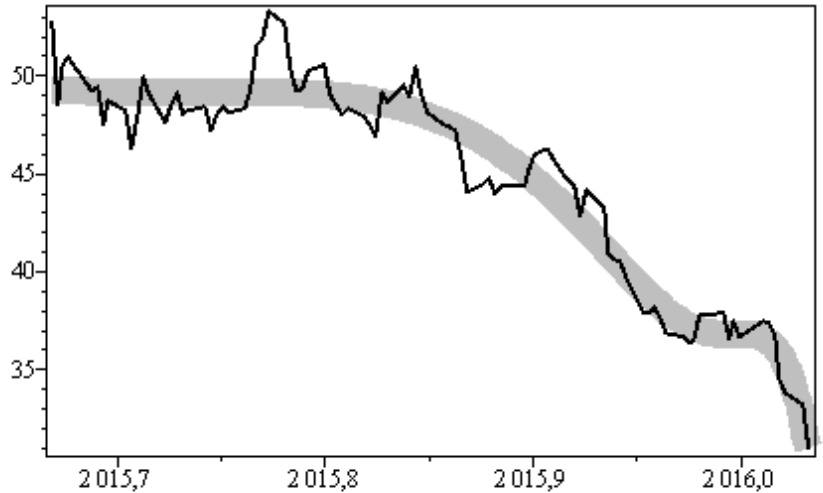

**Fig. 7. Large negative bubble.** *Smooth line corresponds to the following log-periodic parametrization: -24513.408 + 24571.038 (2016.061-t)$^{0.000307}$ + 2.027 (2016.061-t)$^{0.000307}$ cos{8.829 log(2016.061-t)+0.384}. Its point of singularity = 2016.06121 = 22.01.2016 8:34.*

This section will review one more argument, more exactly a system of arguments, in favor of the fact that in the data in Fig. 3 there "sits" a negative bubble. Namely, we will show a system of log-oscillations superimposed over each other which exists in points of Fig. 3 apart from those cited therein and which have approximately the same point of singularity as that in Fig. 3. This property is imminent to different indicators of financial markets, described by (1) (see Sornette 2004). That would additionally indicate that Fig. 3 is, indeed, a negative (sub) bubble.

This, in particular, is supported by an alternative variant of the log-periodic parametrization relative to Fig. 3. It is shown in Fig. 8.

As one can see, the resulting point of singularity is less than the point of singularity at Fig. 3 by only 4 days. For log-periodic parametrizations of various indicators of financial markets, the possibility of alternative log-periodic parameterization is a law, which is a manifestation of the same evolutionary process, having a multi-level, fractal nature (see Sornette 2004). That indicates that log-periodic parameterization at Fig. 3, exactly as log-periodic parameterization at Fig. 8 are two different sides of the same process – the process of negative bubble development.

Moreover, it can be seen in Fig 8 that smaller log-oscillations are superimposed over the larger ones, with approximately the same point of singularity, as in Fig. 8. Fig. 9, in fact, shows the same type of log-oscillations as in Fig. 3. The point of singularity is only five days earlier than in Fig. 3. Log-oscillations of Fig. 9 are much more distinct than in Fig. 3, which indicates that log-periodic parameterization of Fig. 3 is not an artifact and that irregularity of Fig. 3 is associated mainly with the superimposition of the log-oscillations shown there over a larger type of log-oscillations of Fig. 8.

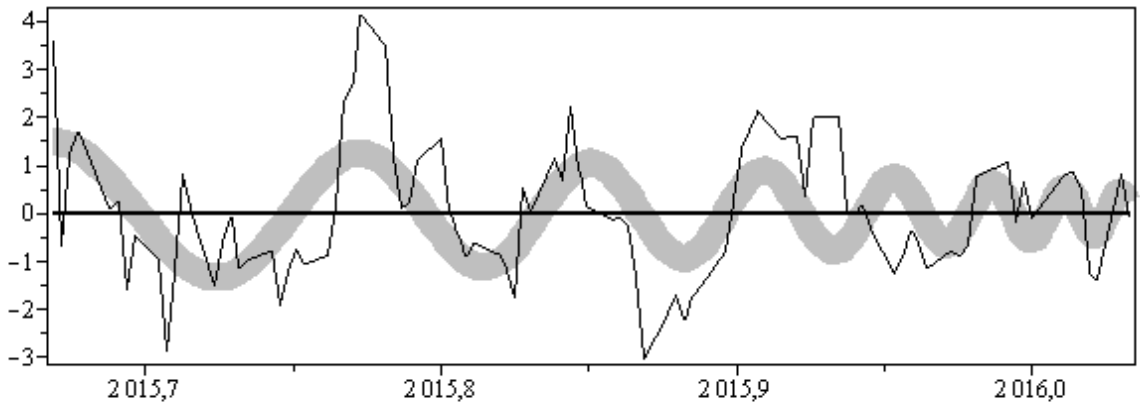

**Fig. 8. Smaller type of log-oscillations superimposed over log-oscillations of Fig. 8** *(the 2$^{nd}$ step of iterations). Here the broken line represents the difference between the broken and smooth lines in Fig. 8. Smooth line corresponds to the following log-periodic parametrization: 2.509 (2016.086-t)$^{0.610}$ cos{50.289 log(2016.086-t)+0.257}, achieved through the minimization of dispersion. Point of singularity = 2016,08578 = 31.01.2016  7:57.*

At the next iteration step one can see another type of log-oscillations as shown in Fig. 10. It is evident that it is this type of log-oscillation pattern that is responsible for the very strong deviation of the broken line from the parametrization in Fig. 3 at a value of time of about 2015.95, which once again indicates that even strong deviations of the broken line from a log-periodic parameterization of Fig. 3 does not mean that there is no negative bubble. After all the processes in the financial markets described by equation (1) always have a complex, multi-layered, "Russian-doll-like" structure, where one type of log-periodic oscillations is superimposed over the 2nd, the 2nd is superimposed over the 3rd, etc. (see Sornette 2004), which can translate into very different, bizarre structures, behind of which it is often not immediately easy to recognize pattern (1).

At Fig. 10 the singularity point has turned out again to be close to the ones obtained above – corresponding to a date that is later than in Fig. 3 by 15 days.

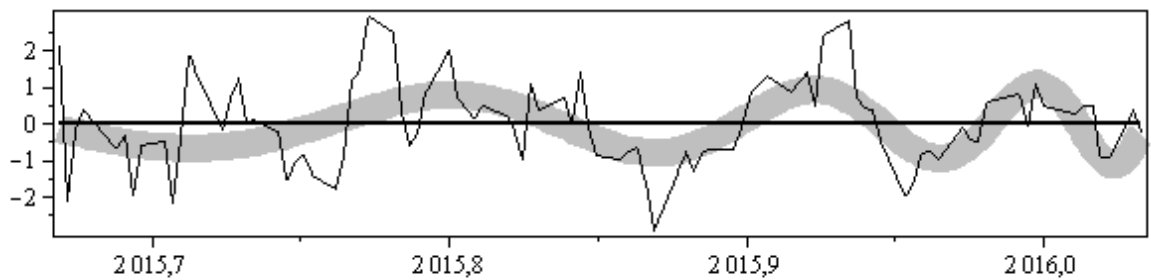

**Fig. 9. Middle type of log-oscillations (3rd step of iterations).** X-axis – same as in **Fig.9.** *Broken line is the difference between the broken line and the smooth line of the Fig. 9. Smooth line is log-parametrization 0.513 cos{29.472 log(2016.117-t)+2.164) / (2016.117-t)$^{0.338}$ with parameters obtained by dispersion minimization. Point of singularity = 2016.11664 = 11.02.2016 14:27.*

I.e. the difference here is larger than that for the previous iteration step from Fig. 9. However, this does not mean some different point of singularity. The error of definition simply does not allow to see the true value of the point of singularity. In Fig. 10 the number of oscillation periods is smaller than in Fig 9; so the accuracy of defining the point of singularity can be lower as well.

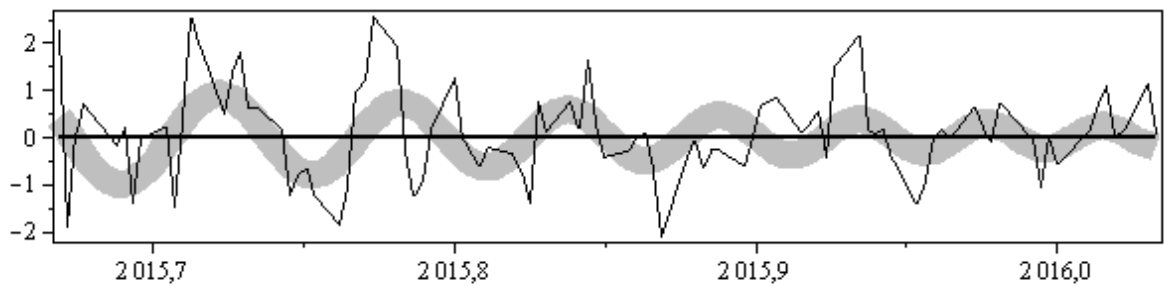

**Fig. 10. The 3rd step of the negative sub-buble iterations.** *Broken line is the difference between the broken line and the smooth line of the Fig. 10. Smooth line is log-parametrization 2.224 (2016.422-t)$^{2.527}$ cos{159.764 log(2016.422-t)+0.334} with parameters obtained by dispersion minimization. Point of singularity = 2016.4218 = 02.06.2016 1:35».*

Iterations can be continued further on. Fig 11 shows the following iteration. Point of singularity again is close to the one obtained in Fig. 3 and exceeds it by 14 days.

**Accuracy assessment of the point of singularity of the negative sub-bubble**

To assess the accuracy of the point of singularity of the negative bubble (Fig. 3) one can variate the point of singularity "manually" within a certain range, while optimizing other parameters at the same time. As a result one will be able to detect such a range of values of the singularity point, at which extremes one could visually notice evident discrepancy of parametrization. This margins of the range can well be identified with the error margins as regards the detection of the singularity point. This operation can be done with all the above discussed iterations, starting from Fig. 3 and ending with Fig. 11. In this case, the point of singularity will be evaluated with the highest precision in the case of the largest number of oscillations. This applies to log-oscillations of the type shown in Fig. 3. Therefore, we will make assessment for this very type.

*Assessment of the lower limit of the singularity zone*

If we gradually decrease the point of singularity and optimize all other parameters, then instead of Fig. 3 we can get a log-periodic parametrization shown in Fig. 12 with visible discrepancies of parametrization.
 The discrepancy is clearly visible only if we consider separately the oscillatory component of the parameterization in a logarithmic time scale (see Fig. 13).

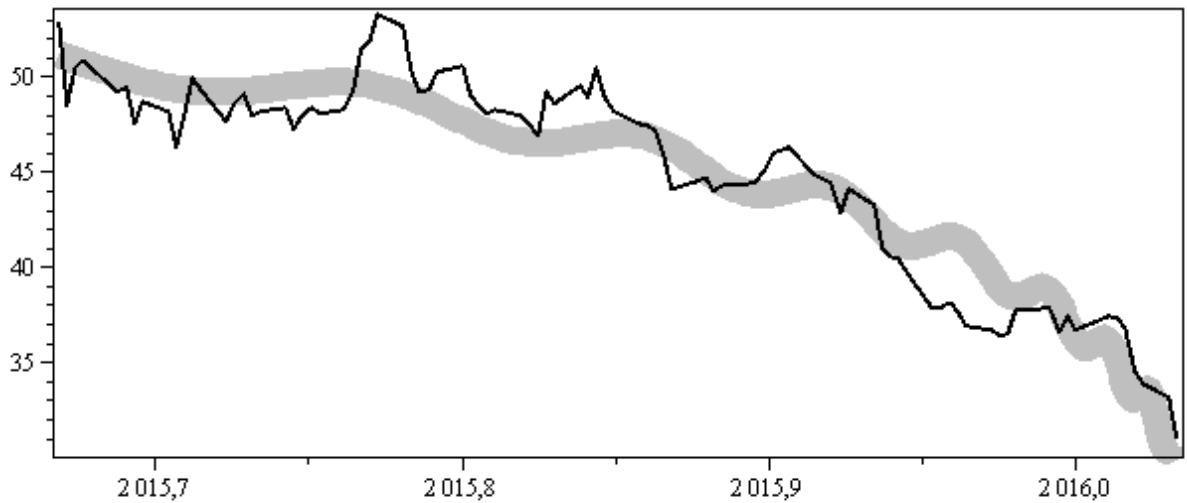

**Fig. 12.** Log-periodic parametrization of the negative sub-bubble at artificially lowered point of singularity**.** *Smooth line: -8.739 + 68.569 $(2016.053-t)^{0.139}$ + 1.157 $(2016.053-t)^{0.139}$ cos(38.022 log(2016.053-t)+1.964). Point of singularity = 2016.053 = 19.01.2016 8:35.*
 In the right part of Fig. 13 there is a clear discordance between the broken line and the parametrization. Thus, the true point of singularity cannot be less than its point of singularity, i.e. the date of 19.01.2016.

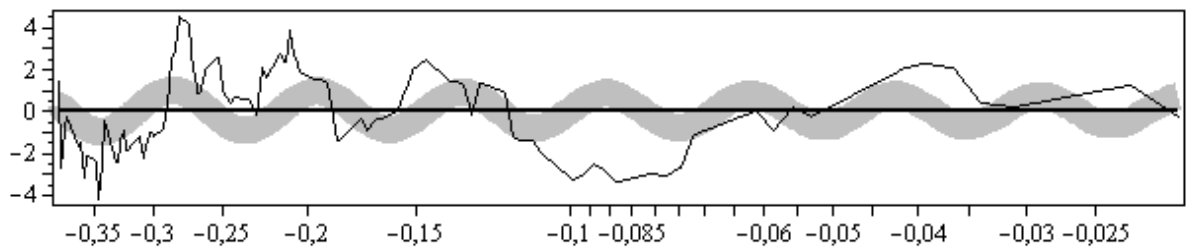

**Fig. 13. Oscillatory component of Fig. 12.** *X-axis is time in decimal, logarithmic scale with the start point at the point of singularity from **Fig. 12**.*

With all the limitations of visual criterion for the degree of "mismatch", the estimated lower limit gives an estimate of the singularity point value before which the real singularity cannot occur with a probability close to 100%. This is due to the fact that in Fig. 13 vibrations of points to the right are approximately in antiphase with the parameterization. Typically, the log-parameterization provides a situation where the extrema of parameterization approximately coincide with the extremes of the data points – as, for example, is the case for Fig. 8 (if on the corresponding iteration step to select a log parameterization which has a minimum dispersion of all the possible parameterizations) or Fig. 10. However, sometimes (but rarely) we may confront something like almost an antiphase (as may be observed in Fig. 11 at about 2015.9). Therefore, if you gradually change the singularity point of log-parameterization (while optimizing the other parameters) and achieve antiphase for at least some fluctuations, the value of the singular point at which it happened will assess the border of singularity zone for which, with almost 100% probability, the real point of singularity cannot get out. In fact, it is such an assessment that was made in Fig. 13.

*Assessment of the upper limit of the point of the singularity zone*

In a similar way, gradually increasing the point of singularity and optimizing all other parameters we can once more get a "picture" of a visible dissonance of the parametrization and the broken line. The dissonance is already visible in **Fig. 14**.

It is more clearly visible for the oscillatory component of this parametrization shown in **Fig. 15.**

Thus, the true point of singularity cannot be more than ≈ 03.02.2016.

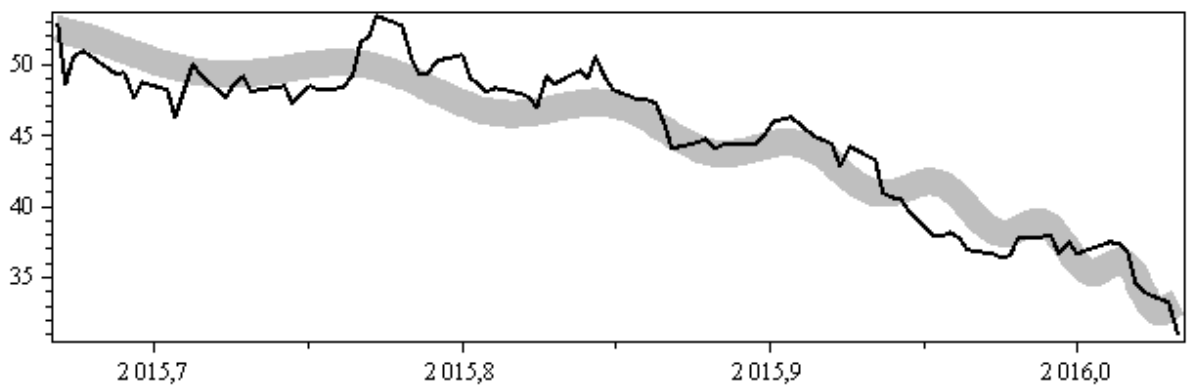

**Fig. 14. Log-periodic parametrization of the negative sub-bubble at artificially lowered point of singularity.** *Smooth line: $-48406.532 + 48466.396\ (2016.093-t)^{0.000201} + 1.015\ (2016.093-t)^{0.000201}\ cos(50.447\ log(2016.093-t)+0.535)$. Point of singularity = 2016.093 = 02.02.2016 23:14:17.*

Thus, the estimate obtained with the method described above shows that the point of singularity of the negative sub-bubble should be situated somewhere between 19.01.2016 and 02.02.2016. In other words, the fall of oil prices corresponding to this negative sub-bubble should end no later than the second half of January 2016. However, we would mention again that afterwards the fall of the oil prices will resume breaking new anti-records until the final collapse of the negative oil price bubble between 24.03.2016 and 15.05.2016.

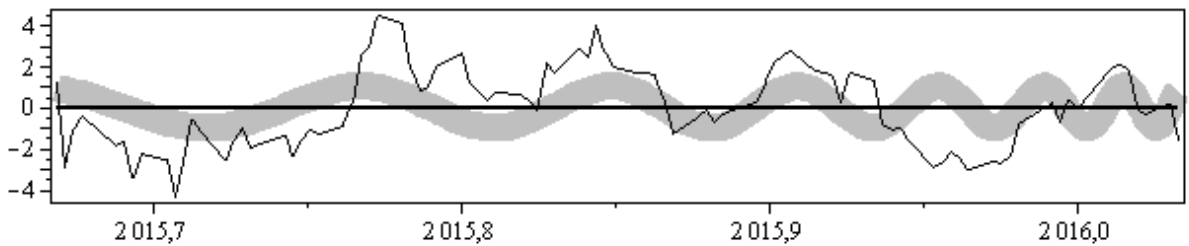

**Fig. 15. Oscillatory component of the Fig 14.** *Broken line is the difference between the broken and main (periodic) components of log-periodic parametrization at Fig. 14 (it is equal to $-48406.532 + 48466.396 (2016.093-t)^{0.000201}$); smooth line is the oscillatory component of the log-periodic parametrization at Fig. 14 (it is equal to $1.015 (2016.093-t)^{0.000201} \cos(50.447 \log(2016.093-t)+0.535))$.*

## Conclusion

Method of log-periodical parametrization of various important indicators in financial markets has wide prospects and is worth using in regular forecasting carried out by governments for country budgets and economic conditions. At the current level of development, it partly turns incapable of providing us with policy-important future values of certain indicators (such as oil price). However, it allows forecasting with policy-sufficient accuracy the critical moments after which the development trends change (e.g., oil price collapse stops). Further development of the method can potentially lead to forecasting certain financial indicators with policy-sufficient accuracy. For this one needs to take into account the fractal character of log-oscillations overlapping each other. Building a fundamental theory of the fractal growth of log-oscillations would allow to solve this task, and government support of the corresponding research could turn out critically important. The new results of such research could start a new chapter in approaches to forecasting the economic (and political) development of particular countries and the world as a whole.

### References

**Akaev A., Fomin A., Tsirel S., Korotayev A. 2010.** Log-Periodic Oscillation Analysis Forecasts the Burst of the "Gold Bubble" in June – August 2011. *Structure & Dynamics* 4/3: 1–11.
**Akaev A., Sadovnichy V., Korotayev A. 2012.** On the dynamics of the world demographic transition and financial-economic crises forecasts. *The European Physical Journal Special Topics* 205/1: 355-373.
**Eia.gov.** http://www.eia.gov/dnav/pet/hist/LeafHandler.ashx?n=PET&s=rbrte&f=D .
**Johansen A., Sornette D., Ledoit O. 1999.** Predicting financial crashes using discrete scale invariance. *Journal of Risk* 1/4: 5–32.

**Johansen A., Sornette D., Wakita H., Tsunogai U., Newman W. I., Saleur H. 1996.** Discrete scaling in earthquake pre-cursory phenomena: Evidence in the Kobe earthquake, Japan. *Journal de Physique I* 6/10: 1391–1402.

**Johansen A., Sornette D. 1999.** Critical Crashes. *Risk* 12/1: 91–94.

**Johansen A., Sornette D. 2001.** Finite-time Singularity in the Dynamics of the World Population and Economic Indices. *Physica A* 294/3–4: 465–502.

**Korotayev A., Tsirel S. 2010.** A Spectral Analysis of World GDP Dynamics: Kondratieff Waves, Kuznets Swings, Juglar and Kitchin Cycles in Global Economic Development, and the 2008–2009 Economic Crisis. *Structure and Dynamics* 4/1: 3–57. URL: http://www.escholarship.org/uc/item/9jv108xp.

**Korotayev A., Zinkina J. 2011*a*.** Egyptian Revolution: A Demographic Structural Analysis. *Entelequia. Revista Interdisciplinar* 13: 139–169.

**Korotayev A., Zinkina J. 2011*b*.** Egyptian Revolution: A Demographic Structural Analysis. *Middle East Studies* 2/5: 57–95.

**Quandl.com.** https://www.quandl.com/data/CHRIS/CME_BZ1-Brent-Crude-Oil-Financial-Futures-Continuous-Contract-1-BZ1-Front-Month .

**Sornette D. 2004.** *Why stock markets crash: critical events in complex financial systems.* Princeton, NJ: Princeton University Press.

**Sornette D., Johansen A. 1997.** Large financial crashes. *Physica A* 245/3–4: 411–422.

**Sornette D., Johansen A. 1998.** A hierarchical model of financial crashes. *Physica A* 261/3–4: 351–358.

**Sornette D., Johansen A. 2001.** Significance of log-periodic precursors to financial crashes. *Quantitative Finance* 1/4: 452–471.

**Sornette D., Sammis C. G. 1995.** Complex critical exponents from renormalization group theory of earthquakes: Implications for earthquake predictions. *Journal de Physique I* 5/5: 607–619.

**Sornette D., Woodard R., Zhou W.-X. 2009.** The 2006–2008 Oil Bubble: evidence of speculation, and prediction. *Physica A* 388: 1571–1576.

**Yan W., Woodard R., Sornette D. 2011.** Diagnosis and Prediction of Market Rebounds in Financial Markets. URL: http://arxiv.org/abs/1003.5926 .